# New Preshower detector for DIRAC Experiment

August 19, 2014

M. Pentia[a,1], S. Aogaki[a], D. Dumitriu[a], D. Fluerasu[a], M. Gugiu[a], V. Yazkov[b]

[a]Horia Hulubei National Institute of Physics and Nuclear Engineering (IFIN-HH),
Reactorului St. 30, P.O.Box MG-6 Măgurele, 077125 Bucharest, Romania

[b]Skobeltsin Institute for Nuclear Physics of Moscow State University, Moscow, Russia

## ABSTRACT

The Preshower (PSh) detector is a component of the DIRAC setup. It is designed to improve rejection efficiency of $e^-e^+$ pairs background in the $\pi\pi$ and $K\pi$ pair measurement. To increase the overall efficiency, a new two-layer structure scintillator Preshower detector has been realized in the region where the Nitrogen Cherenkov detector has been shortened to introduce new detectors. The new Preshower-Cherenkov combination ensures the electron rejection efficiency better than 99.9% in momentum region 1-7 GeV/c.

**Keywords**

Hadron/electron separation; Preshower detector; Electromagnetic shower; Hadronic atoms

## 1. Introduction

In the experimental study [1-6] of $\pi^+\pi^-$ or $K\pi$ hadronic atoms to test low energy QCD [7] by measuring the meson pairs produced by hadronic atoms ionization, there is a significant $e^+e^-$ background. It must be detected and removed to have pure events with meson pairs. Therefore, the electron rejection is an essential problem in the hadronic atoms study and their lifetime measurement.

Principles and functional characteristics of the Preshower (PSh) detector of the DIRAC experiment [8] were presented in [9]. Now, the new PSh detector [10] was redesigned and reconstructed to increase the efficiency of electron rejection especially on the kaon phase space registering.

DIRAC-II setup (see **Figure 1**) configuration [11] was changed relative to the old one by adding two new Cherenkov detectors, one with heavy gas ($C_4F_{10}$) for pion identification and one with aerogel for kaon identification. The old Cherenkov detector (with $N_2$) for electron identification was cut to make room for placement of the two new components. Therefore, reducing by more than half the length of the particle's track, the $N_2$ Cherenkov detector electron rejection efficiency decreased to approx. 85%.

For these reasons, to compensate for lower rejection efficiency, the PSh configuration was changed by adding a new layer to grow the own efficiency.

---

[1] Corresponding author: IFIN-HH, Reactorului St. 30, P.O.Box MG-6 Măgurele, 077125 Bucharest, Romania, E-mail: pentia@cern.ch

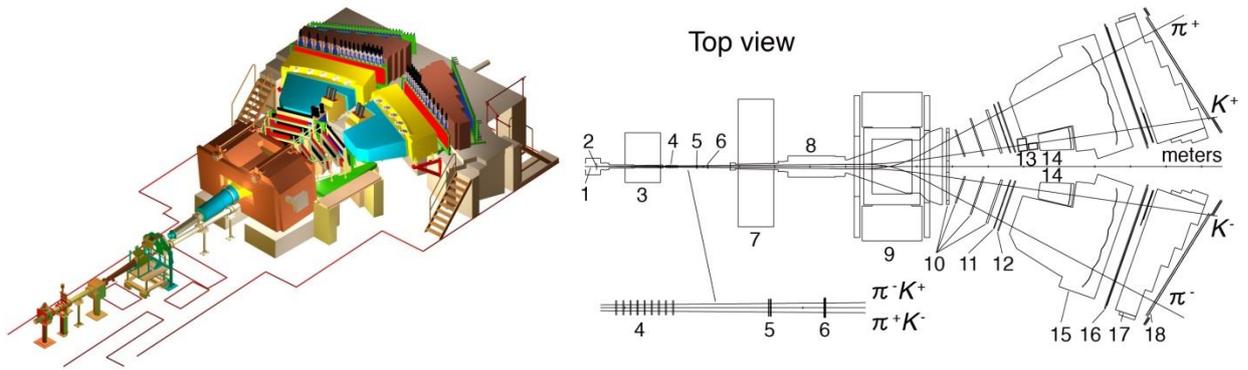

**Figure 1.** DIRAC-II setup.
1. Proton beam; 2. Target station; 3. First shielding; 4. Micro Drift Chambers; 5. Scintillating Fiber Detector; 6. Ionization Hodoscope; 7. Second Shielding; 8. Vacuum Tube; 9. Spectrometer Magnet; 10. Drift Chambers; 11. Vertical Hodoscope; 12. Horizontal Hodoscope; 13. Aerogel Čerenkov; 14. Heavy Gas Čerenkov; 15. Nitrogen Čerenkov; 16. Preshower; 17. Iron absorber; 18. Muon Detector

This way we obtained a new Preshower-Cherenkov ensemble with a reduced spatial extension and an electron rejection efficiency of over 99%.

## 2. Preshower detector configuration within DIRAC-II Experiment

The new PSh detector [10] for DIRAC-II experiment consists of 40 scintillation slabs, placed symmetrically along the two arms, for positive and negative particles (see **Figures 1, 2**).

The PSh geometrical characteristics are presented in **Figure 2.** It contains, in the first layer *Pb* converter of 10*mm* thickness for the first two slabs and 25*mm* for the rest. In the high energy phase space region of the kaon flight, the PSh detector has two layers, to reject both electrons and pion contamination due to Nitrogen Cherenkov detector escape. It contains *Pb* converter of 10*mm* thickness. The detector slabs, placed behind the *Pb* converter, are plastic scintillators BICRON type 408 of 10*mm* thickness.

The main process in the particle discrimination by PSh detector is the detection of the electromagnetic shower produced by electrons. In the PSh configuration the Pb converter thickness is large enough to provide the electromagnetic shower in its primary stage development due to the incident relativistic electrons. But just few percent of incident pions can interact and produce hadronic shower. Therefore, the electromagnetic shower is well-defined in this sector, with a relatively large number of cascade electrons and photons, whereas the pions interact mainly as minimum ionizing particles (mip), without shower production. Electrons are observed by measuring the large amplitude signal produced by electromagnetic shower in Pb converter (4-6 radiation lengths) with the scintillation detector (see **Figure 2**).

In the experimental DIRAC-II configuration the PSh detector works along with the Nitrogen ($N_2$) Cherenkov detector (**Figure 3**), aiming to reject background electron events. The $N_2$ gas radiator is used at normal temperature and pressure. These conditions determine particle momentum threshold values $p^e_{thres} = 20.3 MeV/c$ and $p^\pi_{thres} = 5.5 GeV/c$.

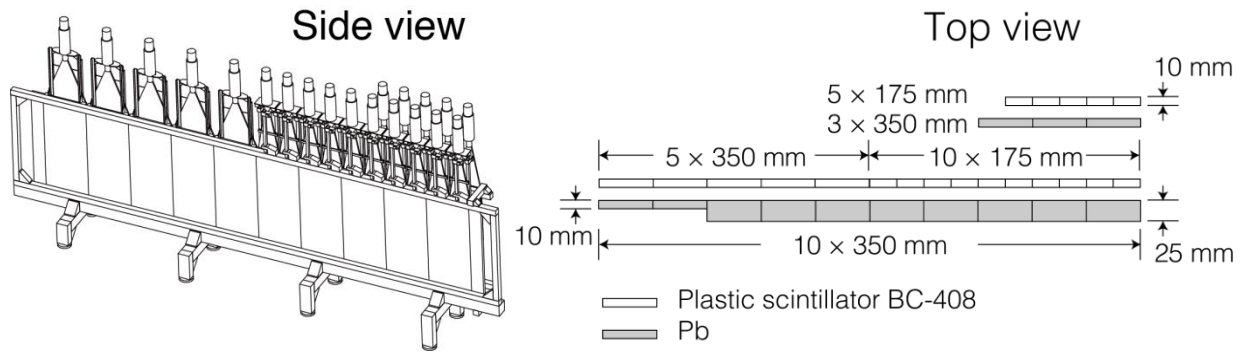

**Figure 2.** Preshower detector (left arm)

The Cherenkov sensitive volume has been reduced in the kaon flight region (see **Figure 4**) to permit insertion of the heavy gas Cherenkov detector for pion identification and aerogel Cherenkov detector for kaon identification. Consequently, the counter length in the pion flight region is 285 cm ("long part") and 140 cm in the kaon flight region ("short part").

The Cherenkov process produces photons in proportion to particle path length in radiator. Therefore the electron detection efficiency in the Kaon region of the $N_2$ Cherenkov detector decreased considerably. Then the *electron rejection efficiency* for "long part" is ~99.5% and for "short part" is ~85%.

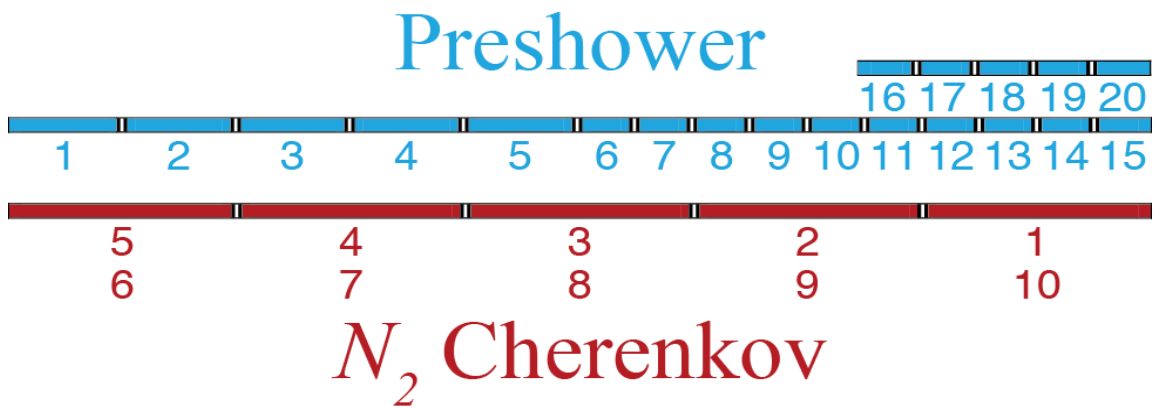

**Figure 3.** Relative positions of the $N_2$ Cherenkov and Preshower elements.
The numbers show the PMT positions (left arm)

Moreover, the high energy pions and accidental coincidences occurring within the trigger time-gate are also labeled as "electron" by Cherenkov signals. The only possibility to reject such wrong „electrons" and increase the overall electron rejection efficiency of the DIRAC-II setup can be done with PSh.

The signal produced in PSh detector by a high energy pion has a low amplitude, because it losses energy mainly by direct ionization as mip's. So the amplitude (ADC) spectrum shows a sharp peak around a low mean value (see **Figure 5**) and a higher amplitude tail due to shower particles produced in *Pb* by some of the pions. The tail contribution increases with the pion energy.

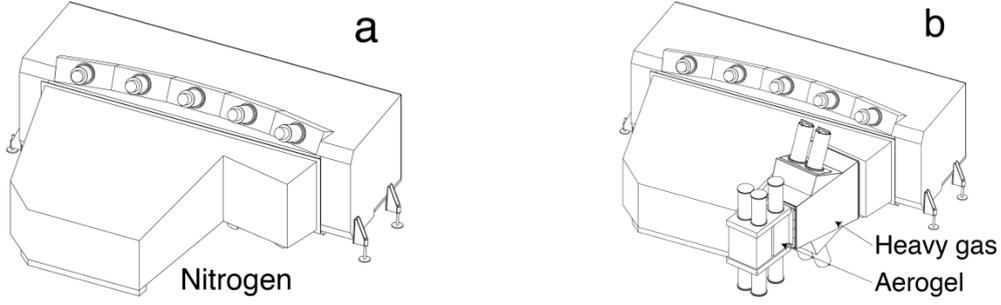

**Figure 4.** Cherenkov Detectors

On the other hand, a relativistic electron incident on the *Pb* converter produces an electromagnetic shower of electrons and photons. In interaction with the scintillation material they produce a large amplitude signal, proportional to the total absorbed energy. Therefore, the amplitude spectrum of high energy electrons shows a broad amplitude distribution (see **Figure 5**) with higher mean value. The pion peak is practically independent on energy (as minimum ionizing particle), but the electron distribution is moving to larger amplitudes for higher energies.

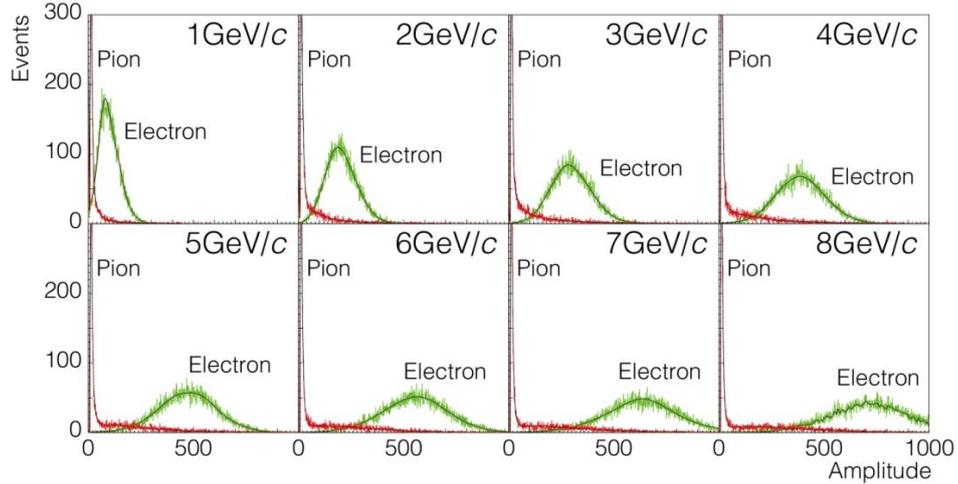

**Figure 5.** The Monte-Carlo simulation of pion and electron amplitude distribution at $5X_0$ depth in Pb converter, for incident energy 1-8 GeV. Scintillator thickness $w_{Sc}=1$ cm.

The PSh electron and pion amplitude spectra are taken in coincidence and anticoincidence with $N_2$ Cherenkov signals. These spectra are partially overlapped, see **Figure 6.** The analysis of the particle spectra assumes the possibility to separate them based on particle type and momentum value.

The separation cut must to be established to evaluate the electron rejection (ratio of the cut right side events and the total number of events in the electron spectrum) and the pion loss (ratio of the cut right side events and the total number of events in the pion spectrum).

The trigger used for PSh ADC spectra aquisition includes vertical (V) and horizontal (H) hodoscopes along $N_2$ Cherenkov (Ch) signals,

- $(V_1 \cdot H_1 \cdot Ch_1)$ x $(V_2 \cdot H_2 \cdot Ch_2)$ for $e^+e^-$ pairs
- $(V_1 \cdot H_1 \cdot \overline{Ch_1})$ x $(V_2 \cdot H_2 \cdot \overline{Ch_2})$ for $\pi^+\pi^-$ pairs

**Figure 6** presents typical electron and pion amplitude spectra, taken with $e^+e^-$ and $\pi^+\pi^-$ triggers.

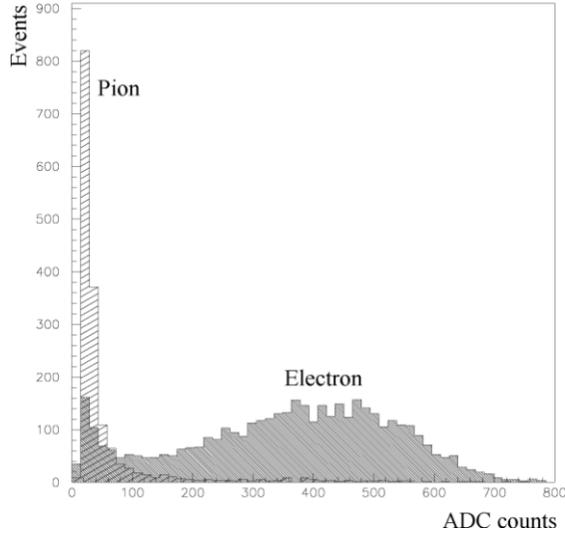
**Figure 6.** Preshower pion and electron spectra

## 3. Preshower detector study by particle tracking

For PSh study, the data analysis program selected events with only one track found by Drift Chamber (DC) of each arm. Tracks were extended to the plane of PSh detector to define the expected coordinate hit position and corresponding slab number. Each track was characterised by the straight line parameters and by particle momentum. As a result of the charged particle track deviation by DIRAC spectrometer magnet the low momentum (1-2 GeV/c) particles were hitting the outermost PSh slabs, while the high momentum (5-6 GeV/c) particles hit the innermost PSh slabs (see **Figure 1**).

**Figure 7** presents the (x,y)-scatter plot of all DC reconstructed tracks detected by the PSh detector (see also **Table 1** column 3).

The x axis of the PSh local coordinate system was horizontally aligned, along the PSh basis, originating in the center of the PSh plane, directed to the right side (see **Figures 7, 8, 9**).

In the DIRAC-II setup configuration, the PSh detector doesn't was included in the DIRAC trigger system. It simplifies measurement of the detector efficiency.

Analysing the DC reconstructed tracks, we saw not all the tracks coming to PSh are detected with proper signals in the tracked slab. Some of them are deviated by multiple scattering in materials between Drift Chambers and PSh detector.

**Figure 8** presents the electron detection efficiency (ratio of the detected tracks in the proper slab over the DC reconstructed tracks) as a function of x-coordinates of the DC tracks in the PSh detector planes.

**Figure 9** presents the (x,y)-scatter plot of electron tracks do not detected by the tracked slab (see also **Table 1** column 4). The interspaces between slabs show a lower efficiency, with electrons scattered to neighbor slabs. So individual slabs can be identified. The other strips are produced by the scattered electrons in the thick materials before PSh as $N_2$ Cherenkov mirror support (central horizontal line) and $N_2$ Cherenkov detector wall (intensive vertical lines at $x \approx \pm 80$cm).

The x-dependence of detection efficiency shows an enhancement of nondetected tracks in proper slab with strips along slab boundaries as deviated tracks toward adjacent slabs.

The method can be used as a radiography imaging technique for materials before PSh detector.

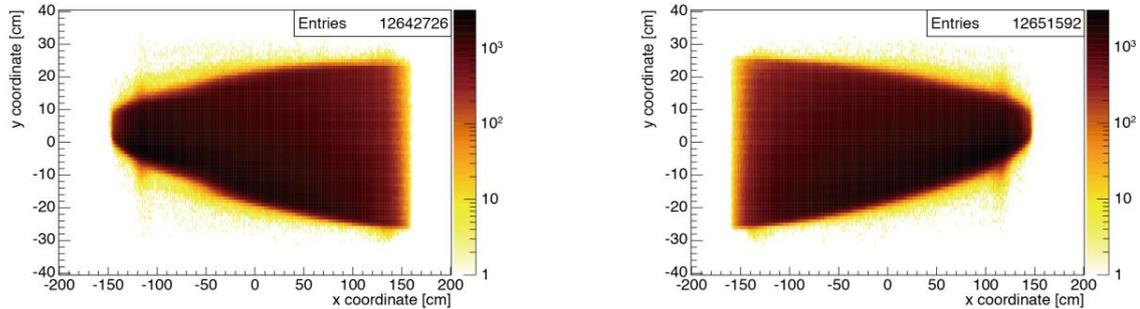

**Figure 7.** Coordinates of electron tracks **detected** (**signals**) by the tracked slab.

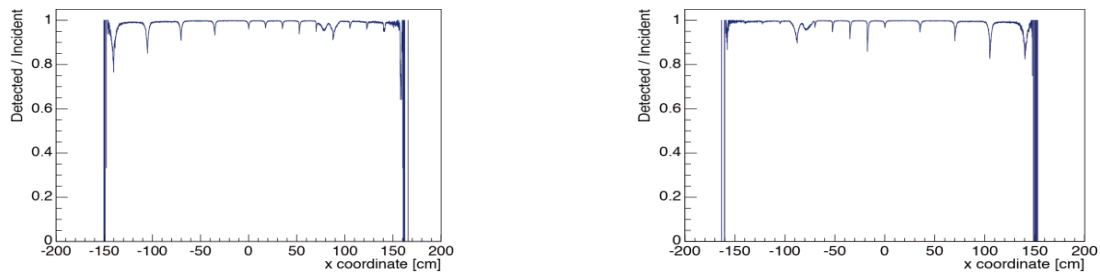

**Figure 8.** Electron (positron) detection **efficiency** by tracked slab as a function of x-coordinate of the DC tracks on PSh detector planes

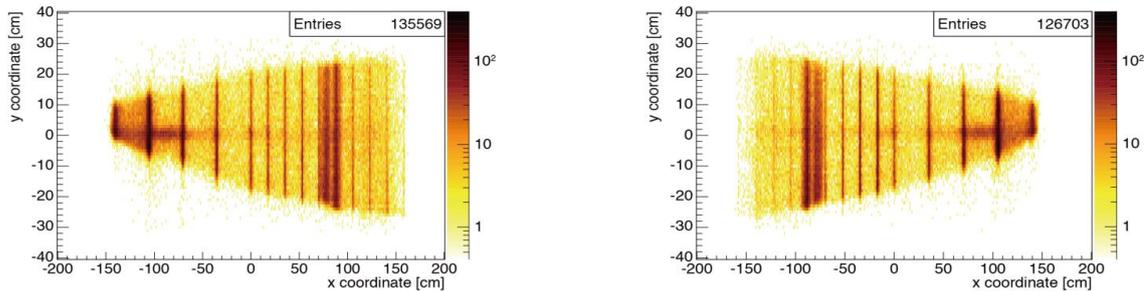

**Figure 9.** Coordinates of electron tracks **not detected** (**no signal**) by tracked slab. Vertical lines show the PSh slab edges, with electrons scattered to neighbor slabs. The other strips are produced by the scattered electrons in thick materials before PSh as $N_2$ Cherenkov detector mirror support (central horizontal line) and $N_2$ Cherenkov detector wall (intensive vertical lines at $x \approx \pm 80$cm).

**Table 1.** Preshower track and signal distribution

| Arm (1) | Tracks on PSh (2) | Signals on tracked slab (3) | No signal on tracked slab (4)=(2)-(3) | No Signal on tracked slab but on other one (5) | No signal on any slab (6)=(4)-(5) |
|---|---|---|---|---|---|
| Left | 12,778,295 (100%) | 12,642,726 (98.94%) | 135,569 (1.06%) | 123,181 (0.964%) | 12,388 (0.097%) |
| Right | 12,778,295 (100%) | 12,651,592 (99.01%) | 126,703 (0.99%) | 112,231 (0.878%) | 14,472 (0.113%) |

## 4. Preshower electron rejection and pion detection efficiencies

The PSh amplitude spectra measurements [12] along the DIRAC setup with the pion and electron trigger, have been registered for every PSh detector slab. Typical experimental spectra, for pions and electrons for some intermediate energies, slab number 4, 12 and 14 of the right arm (negative particles), are presented in **Figures 10, 11, 12**. They are registered respectively in anticoincidence and coincidence with the $N_2$ Cherenkov detector signals. In can be seen that as the energy increases (higher slab number), the pions are increasingly more present in electron spectra.

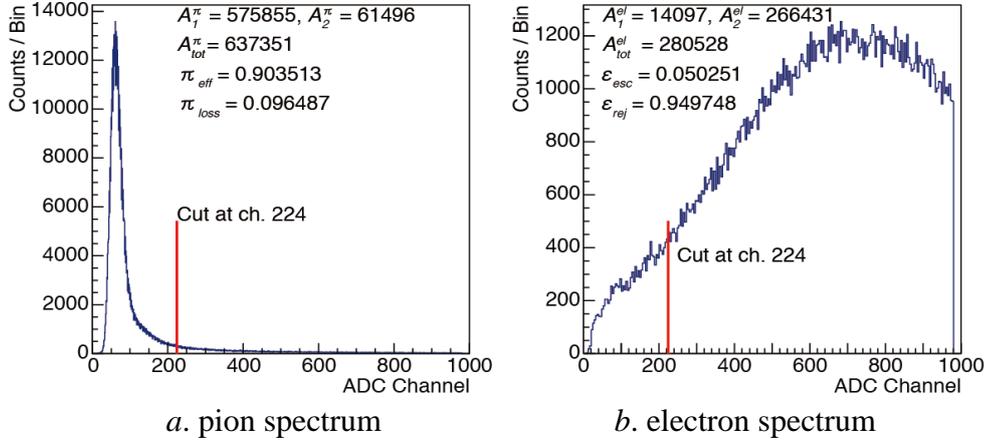

*a*. pion spectrum      *b*. electron spectrum
**Figure 10.** Preshower pion and electron spectra, slab 4, right arm

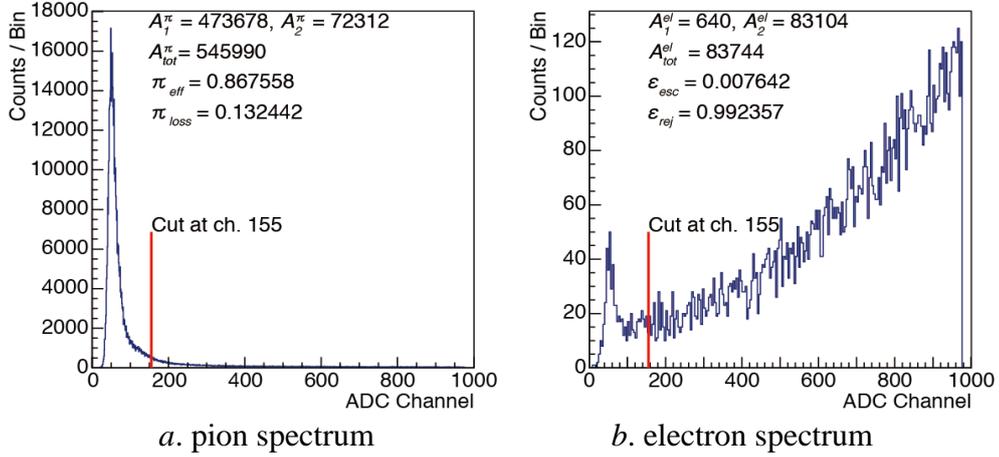

*a*. pion spectrum      *b*. electron spectrum
**Figure 11.** Preshower pion and electron spectra, slab 12, right arm

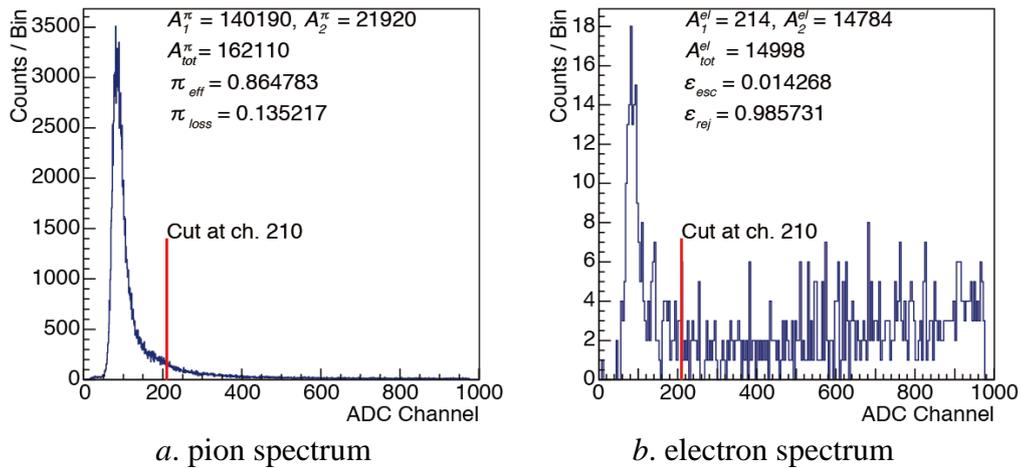

*a*. pion spectrum      *b*. electron spectrum
**Figure 12.** Preshower pion and electron spectra, slab 14, right arm

The cut channel between the pion and electron amplitude distributions is used to determine the number of events $A_1$ and $A_2$ ($A_{tot}=A_1+A_2$), and so to define, in the pion spectra, the *pion detection efficiency* (see **Figure 13**)

$$\pi_{eff} = A_1^\pi / A_{tot}^\pi \quad (1)$$

and the *pion loss*

$$\pi_{loss} = A_2^\pi / A_{tot}^\pi \quad (2)$$

as well as in the electron spectra, the *electron escape*

$$\varepsilon_{esc} = A_1^{el} / A_{tot}^{el} \quad (3)$$

and *electron rejection efficiency* (see **Figure 14**)

$$\varepsilon_{rej} = A_2^{el} / A_{tot}^{el} \quad (4)$$

The cut channel position has been determined by a Gaussian fit of the pion peak, and was placed at the 7σ distance to the right.

As the energy increases (higher slab number), some pion signals are present also in the electron spectra (see **Figures 11-12**) due to higher energy pions, which produce Cherenkov radiation. Therefore the Cherenkov detector cannot separate efficiently high energy electrons and pions. In the PSh electron spectra, they produce a pion contamination. Therefore the analysis of the PSh electron spectra can reject the pion signals and finally obtain a higher overall electron rejection for DIRAC experiment (**Figure 14**). The lower electron rejection for the PSh edge slabs (15 and 20) is due to partial coverage with particle flux. The electron rejection efficiency for one PSh layer, $\varepsilon_{rej}$ is greater than 90%.

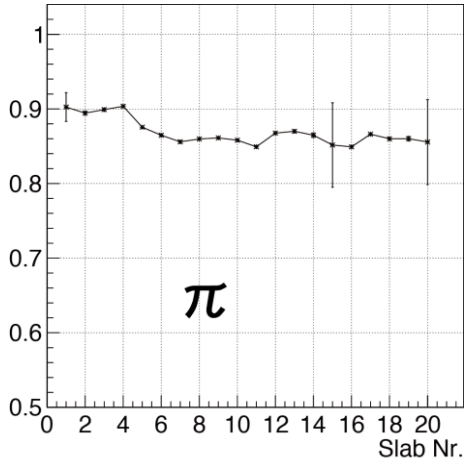
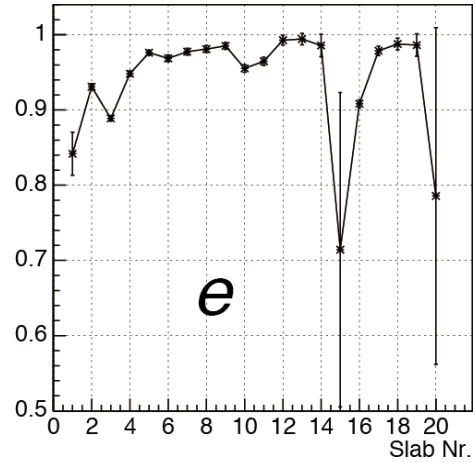

**Figure 13.** One-layer preshower pion detection efficiency (right arm)

**Figure 14.** One-layer preshower electron rejection efficiency (right arm)

### 4.1 Two-layer preshower electron rejection and pion detection efficiencies

To increase the PSh efficiency, a new layer has been added in the region of kaon phase space, where the Cherenkov efficiency is lower. This second layer will detect the pions and electrons that cannot be detected or escaped first layer. So, the second PSh layer will process the high amplitude pions (higher than cut level) and the low amplitude electrons (lower than cut level).

**Figures 15, 16** show the pion and electron spectra for the slab pairs 12+17 on the right arm. For particle selection with signal in both the first and the second layer, we used additional equal momenta condition ($p_{11}=p_{16}$, $p_{12}=p_{17}$, $p_{13}=p_{18}$, etc.) of the DC tracks, for all detected pairs.

In **Figure 15**, the first pion spectrum (slab 12) are produced by the I-st layer, and the second pion spectrum (slab 17) are produced by the lost pions in the I-st layer and detected by the II-nd layer. Now the overall *pion detection efficiency* is:

$$\pi_{eff} = \pi_{eff\text{-}I} + \pi_{loss\text{-}I} * \pi_{eff\text{-}II} \qquad (5)$$

The two-layer *pion detection efficiency* $\pi_{eff}$ values for the slab pairs 1=(11+16), 2=(12+17), 3=(13+18) and 4=(14+19) have been evaluated and plotted in **Figure 17**. The outermost pairs (15+20) for both arms have not been used due to partial covering of the corresponding detector surface (see **Figure 9**). They have not been included in the analysis.

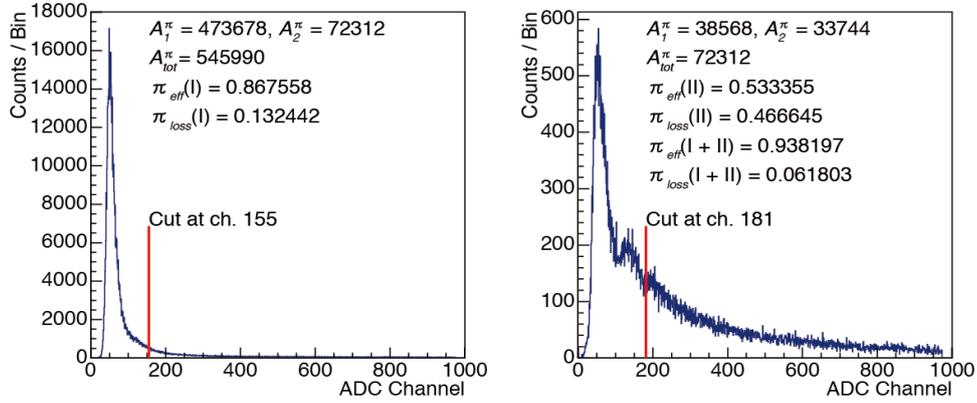

**Figure 15.** Two-layer PSh pion spectra, slab 12 & 17 (right arm)

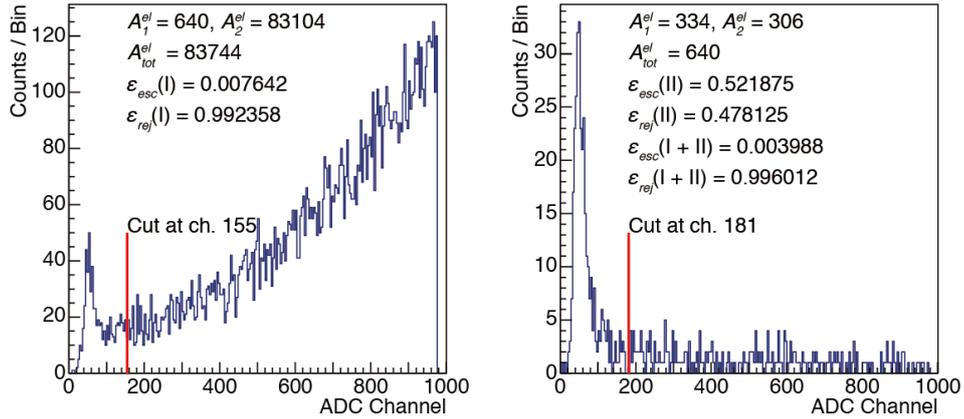

**Figure 16.** Two-layer PSh electron spectra, slab 12 & 17 (right arm)

Similarly in the **Figures 16**, the first electron spectrum (slab 12) are produced by the I-st layer, and the second electron spectrum (slab 17) are produced by the escaped electrons in the I-st layer and detected by the II-nd layer. Now the overall *electron rejection efficiency* is:

$$\varepsilon_{rej} = \varepsilon_{rej\text{-}I} + \varepsilon_{esc\text{-}I} * \varepsilon_{rej\text{-}II} \qquad (6)$$

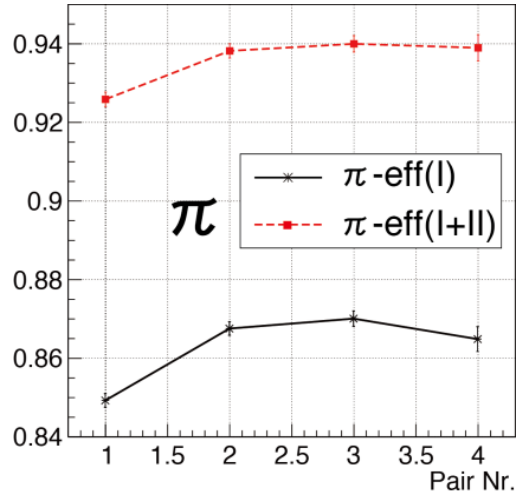 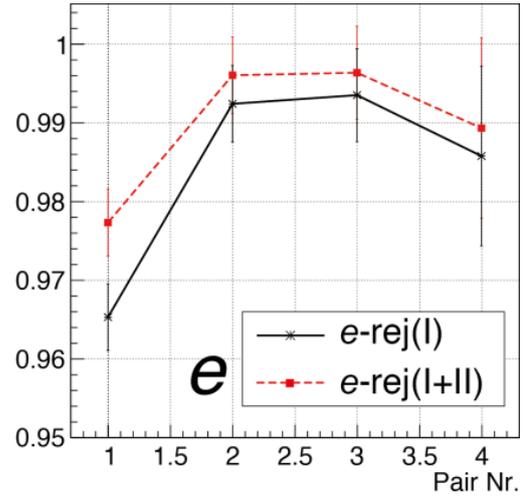

**Figure 17.** Two-layer preshower pion detection efficiency (right arm)

**Figure 18.** Two-layer preshower electron rejection efficiency (right arm)

The two-layer *electron rejection efficiency* $\varepsilon_{rej}$ for the slab pairs 1=(11+16), 2=(12+17), 3=(13+18) and 4=(14+19) have been evaluated and plotted (see **Figure 18**). The outermost pairs (15+20) for both arms have not been used due to partial covering of the corresponding detector surface (see **Figure 9**). They have not been included in the analysis.

## 5. Electron rejection by PSh detector in DIRAC experiment

The DIRAC experiment identifies $\pi^+\pi^-$ (or $\pi K$) hadronic atom by measuring the relative momentum $Q$ of the two components after ionization. Therefore, the background electron-positron pairs must be once more rejected after $N_2$ Cherenkov rejection. The use of PSh information can do it. In **Figure 19** it is presented the PSh electron rejection and pion detection efficiencies within transversal relative momentum $Q_T$ distributions.

The original distributions (without PSh criteria) are drawn by solid line. Experimental $Q_T$ distribution of e$^+$e$^-$ pairs shows a narrow contribution (<1.5 MeV/c), wrongly identified as e$^+$e$^-$ pairs, in **Figure 19** with solid line. As a result, the e$^+$e$^-$ background induces an essential bias in the fit procedure. Strong rejection of this background is needed.

Direct rejection of e$^+$e$^-$ pairs with only the cut criterion on amplitudes of each particle in PSh detector leads to essential contamination of $\pi^+\pi^-$ pairs with electron escape (2). To prevent it, an additional algorithm of e$^+$e$^-$ background rejection has been developed to subtract electron admixture in the pion region.

After PSh cut electron rejection the $Q_T$ distribution in **Figure 19** are drawn by dash line, and after additional subtraction of electron admixture in the PSh cut data, they are drawn by dots.

After PSh cut, the electron rejection is 87.5% and after additional electron admixture subtraction, it is 99.99%. This was accompanied by a small $\pi^+\pi^-$ pair loss. So, in the first analysis the pion detection efficiency is 97.8% and in the second analysis it is 97.5%. So, the pion loss is 2.5%.

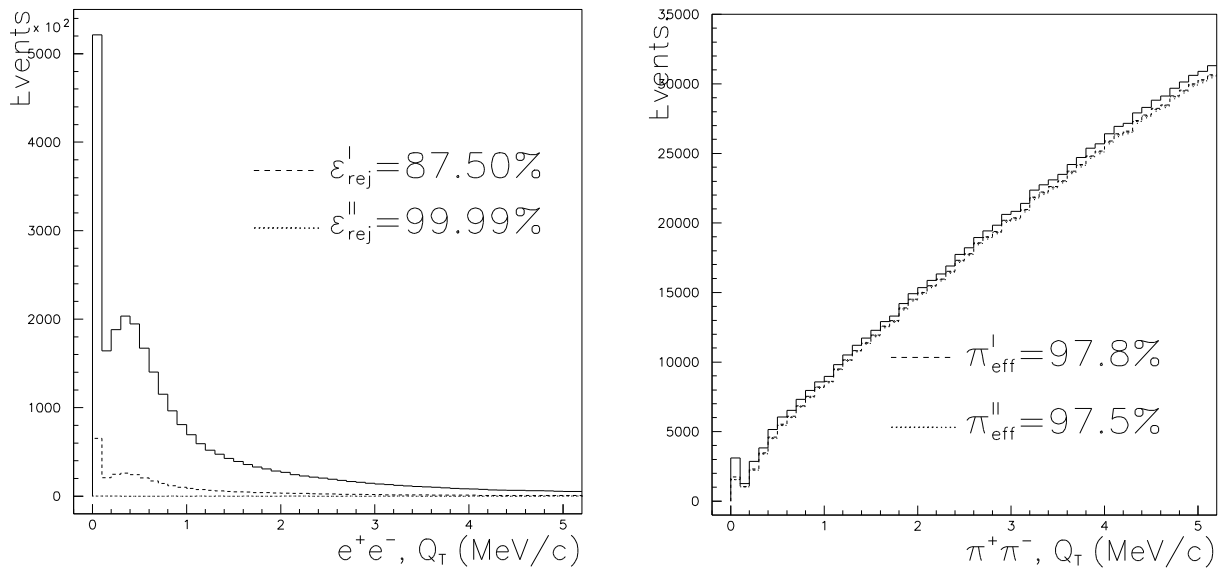

**Figure 19.** Distributions of the transversal relative momentum $Q_T$ of $e^+e^-$ and $\pi^+\pi^-$ pairs without PSh conditions (continuous line) and, I. after PSh cut electron rejection (dash line) and II. after additional subtraction of electron admixture (dots) in the PSh cut data.

**Acknowledgements**

The authors are indebted to members of the DIRAC Collaboration for fruitful discussions and technical support, in particular L. Nemenov, J. Schacher and D. Drijard. This work was supported by Ministry of Education and Research under project CAPACITATI M.III Contract 5/2012 (Romania), the Ministry of Education and Science and RFBR grant 01-02-17756-1 (Russia)